\documentclass{article}%
\usepackage{amsmath}
\usepackage{amsfonts}
\usepackage{amssymb}
\usepackage{graphicx}%
\newtheorem{theorem}{Theorem}
\newtheorem{acknowledgement}[theorem]{Acknowledgement}

\newtheorem{definition}[theorem]{Definition}

\newtheorem{proposition}[theorem]{Proposition}

\begin{document}

\title{Coherent states of a particle in a magnetic field and the Stieltjes moment problem}
\author{J.P. Gazeau\thanks{Laboratoire APC, Universit\'e Paris Diderot \textit{Paris
7} 10, rue A. Domon et L. Duquet 75205 Paris Cedex 13, France, E-mail:
gazeau@apc.univ-paris7.fr}, M.C. Baldiotti\thanks{E-mail:
baldiott@fma.if.usp.br}, and D.M. Gitman\thanks{E-mail: gitman@dfn.if.usp.br}\\Instituto de F\'{\i}sica, Universidade de S\~{a}o Paulo,\\Caixa Postal 66318-CEP, 05315-970 S\~{a}o Paulo, S.P., Brazil}
\maketitle

\begin{abstract}
A solution to a version of the Stieltjes moment problem is presented. Using
this solution, we construct a family of coherent states of a charged particle
in a uniform magnetic field. We prove that these states form an overcomplete
set that is normalized and resolves the unity. By the help of these coherent
states we construct the Fock-Bergmann representation related to the particle
quantization. This quantization procedure takes into account a circle topology
of the classical motion.

\bigskip


\smallskip


\end{abstract}

\section{Introduction}

Constructing semiclassical or ``classical-like'' states in quantum mechanics
is in general an open problem. For systems with quadratic Hamiltonians there
exists a well-known procedure to construct the so-called coherent states (CS,
or Glauber-Klauder-Sudarshan, or standard CS), which usually are accepted as
quantum states that behave like their classical counterpart, see e.g.
\cite{ab1,ab2,ab3,ab5}. The CS are widely and fruitfully being utilized in
different areas of theoretical physics. The standard CS turned out to be
orbits of the Heisenberg-Weyl group. This observation allowed one to formulate
by analogy some general definition of CS for any Lie group
\cite{ab4,aag,ab6,ab7} as orbits of the group factorized with respect to a
stationary subgroup. There exists a connection between the CS and the
quantization of classical systems, in particular, systems with a curved phase
space, see e.g. \cite{ab8,ab9}.

In \cite{KowRe05} a modified approach to constructing semiclassical or
coherent states (we call them also CS) was proposed. A technical realization
of the approach recipes depends on each concrete case, in particular, a
principal one is the problem of proving the resolution of the unity by the
constructed CS. In the present article, we construct coherent states for a
charged particle in a constant and uniform magnetic field closely by following
the approach of \cite{KowRe05}, more exactly a squeezed version of these
states. By solving a specific version of the Stieltjes moment problem we make
explicit the resolution of the unity of the constructed CS and so become able
to perform the Berezin-Klauder-Toeplitz or, more simply CS, quantization of
the complex plane. A generalization of the obtained results to a model on a
non-commutative plane will be the topic of a further work.

As far as physical applications are concerned, the resolution of the unity by
CS is fundamental for the analysis, or decomposition, of states in the Hilbert
space of the problem, or of operators acting on this space. In particular, it
allows for a \textquotedblleft classical\textquotedblright\ reading of quantum
dynamical systems, in Schr\"{o}dinger representation, through the time
behavior of mean values of quantum observables in coherent states. Nice
illustrations of this approach are provided by Perelomov in \cite{ab4}. It was
precisely this symbolic formulation that enabled Glauber and others to treat a
quantized boson or fermion field like a classical field, particularly for
computing correlation functions or other quantities of statistical physics,
such as partition functions and derived quantities.

\section{Coherent states of a particle in magnetic field}

Consider a charged particle with charge $e$ and mass $\mu$ placed in a uniform
and constant magnetic field of magnitude $B$ in the $z$-direction. The motion
of the particle in a plane perpendicular to the magnetic field can be
described by the quantum Hamiltonian ($c=\hbar=1$)%
\begin{equation}
H=\frac{1}{2\mu}\left(  P_{1}^{2}+P_{2}^{2}\right)  ~,\ P_{i}=p_{i}%
-eA_{i}~,\ A_{i}=-\frac{B}{2}\varepsilon_{ij}x^{j}~,\ i,j=1,2, \label{A.0}%
\end{equation}
where $x^{i}$ and $p_{i}$ are canonical operators of coordinates and momenta
of the particle and $\varepsilon_{ij}~\left(  \varepsilon_{12}=1\right)  $\ is
the Levi-Civita tensor. It is useful to introduce operators $x_{0}^{i}$, which
are integrals of motion and correspond to the orbit center coordinates,
\[
x_{0}^{i}=x^{i}+\frac{1}{\mu\omega}\varepsilon^{ij}P_{j}~,\ \omega=\frac
{eB}{\mu}~,
\]
and also the angular momentum operator of the relative motion $J$, which in
the present case is just proportional to the Hamiltonian,
\[
J=-\frac{1}{\omega}H=\frac{1}{2}\left(  r^{1}P_{2}-r^{2}P_{1}\right)
~,\ r^{i}=x^{i}-x_{0}^{i}~.
\]
Two independent Weyl-Heisenberg algebras underlie the symmetries and the
integrability of the model. The first one concerns the operators $r_{0\pm
}=x_{0}^{1}\pm ix_{0}^{2}$ that obey $[r_{0-},r_{0+}]=2/\mu\omega$. The second
one concerns the relative motion operators $r^{i}$, $r_{\pm}=r^{1}\pm
ir^{2}=-\left[  P_{2}\mp iP_{1}\right]  /\mu\omega$ with $[r_{+},r_{-}%
]=2/\mu\omega$. They allow one to construct a Fock space with orthonormal
basis $\{|m,n\rangle\}$ obtained by repeated actions of the normalized raising
operators:%
\[
\sqrt{\mu\omega/2}r_{0+}|m,n\rangle=\sqrt{m+1}|m+1,n\rangle~,\ \sqrt{\mu
\omega/2}r_{-}|m,n\rangle=\sqrt{n+1}|m,n+1\rangle~.
\]
Like in \cite{KowRe05}, the CS $\left\vert z_{0},\zeta\right\rangle $ are
introduced as solutions of the eigenvalue problems
\begin{equation}
r_{0-}\left\vert z_{0},\zeta\right\rangle =z_{0}\left\vert z_{0}%
,\zeta\right\rangle ~,\ Z\left\vert z_{0},\zeta\right\rangle =\zeta\left\vert
z_{0},\zeta\right\rangle ~,\ z_{0},\zeta\in\mathbb{C}~, \label{A.1}%
\end{equation}
with $Z=e^{-J+\frac{1}{2}}r_{+}$. The commutation relations $\left[  J,r_{\pm
}\right]  =\pm r_{\pm}$ reproduce the appropriate\ algebra to study the
circular motion, see \cite{KowReP96}. These normalized CS are tensor product
of the state $\left\vert \zeta\right\rangle $, that is an eigenvector of $Z$
with the standard CS $\left\vert z_{0}\right\rangle $. They read in terms of
the Fock basis,
\begin{equation}
|z_{0},\zeta,\rangle=\frac{1}{\sqrt{\mathcal{N}(z_{0},\zeta)}}\sum
_{m,n=0}^{\infty}\left(  \frac{\mu\omega}{2}\right)  ^{\frac{m+n}{2}}%
\,\frac{z_{0}^{m}}{\sqrt{m!}}\,\frac{\zeta^{n}\,e^{-\frac{1}{2}(n+\frac{1}%
{2})^{2}}}{\sqrt{n!}}\,|m,n\rangle\,, \label{CSKR}%
\end{equation}
where $\mathcal{N}$ stands for a normalization factor.

One can easily see that the time evolution of the CS states is only reduced to
the classical time evolution of the parameter $\zeta$. Indeed, using the
relations $\left[  r_{0\pm},H\right]  =0$ and $e^{itH}Ze^{-itH}=e^{-i\omega
t}Z$, and applying the evolution operator $U\left(  t\right)  =\exp\left(
-iHt\right)  $ to the state $\left\vert z_{0},\zeta\right\rangle $ we obtain%
\[
U\left(  t\right)  \left\vert z_{0},\zeta\right\rangle =\left\vert z_{0}%
,\zeta\left(  t\right)  \right\rangle ,\ \zeta\left(  t\right)  =\zeta
\exp\left(  -i\omega t\right)  ~.
\]

In order to play their role of \textquotedblleft
classical-between-quantal\textquotedblright\ bridge, the states $\left\vert
z_{0},\zeta\right\rangle $\ have to resolve the identity in the above
Fock-Hilbert space. This specific (over-)completeness can be proved by
resolving the identity in the corresponding Fock-Bergmann representation. As
was stated in \cite{KowRe05}, the task is equivalent to solving the following
moment problem%
\begin{equation}
\int_{0}^{\infty}t^{n}\varpi_{q}\left(  t\right)  \,dt=n!q^{\dfrac{n\left(
n+1\right)  }{2}}\,,\ q=e^{2}\,,\label{A.2}%
\end{equation}
for an unknown weight function $\varpi\left(  t\right)  $. Let us generalize
the above problem and, consequently, obtain a squeezed version of the CS
(\ref{A.1}), by introducing the following displacement operator%
\begin{equation}
Z_{\lambda}=\exp\left[  \frac{\lambda}{2}\left(  \frac{1}{2}-J\right)
\right]  r_{+}~.\label{A.3}%
\end{equation}
This operator coincides with $Z$ from (\ref{A.1}) for $\lambda=2$, and with
just $r_{+}$ for $\lambda=0$ (or $q=1$), i.e., the case where we have the
tensor product of standard coherent states, called in this context the
Malkin-Man'ko CS \cite{MalMa68}. For an arbitrary $\lambda$ the operator
$Z_{\lambda}$ controls the dispersion relations of the angular moment and of
the position operators. In this case, the construction of the resolution of
identity from the eigenstates of the above operator, and consequently the
proof that they form an (over-)complete set, is equivalent to solving the
moment problem of the form%
\[
\int_{0}^{\infty}t^{n}\varpi_{q}\left(  t\right)  \,dt=n!\,q^{\dfrac{n\left(
n+1\right)  }{2}}\,,\ q\equiv e^{\lambda}~,\ \lambda\geq0~,
\]
for some unknown weight function $\varpi_{q}\left(  t\right)  $. Below, we
find $\varpi_{q}$ for an arbitrary $q\in\lbrack1,\infty)$ and deal with the
corresponding CS as the eigenstates of the operator $Z_{\lambda}$.

\section{Solving Stieltjes moment problem}

Let us consider the classical phase space $\mathbb{C}^{2}=\{\mathbf{x}=\left(
z,\zeta\right)  \,,\,z\in\mathbb{C}\,,\,\zeta\in\mathbb{C\}}$ provided with
the measure:%
\begin{equation}
\mu\left(  d\mathbf{x}\right)  =e^{-\left\vert z\right\vert ^{2}}\,\frac
{d^{2}z}{\pi}\,\varpi_{q}\left(  \left\vert \zeta\right\vert ^{2}\right)
\,\frac{d^{2}\zeta}{\pi}\,, \label{measphsp}%
\end{equation}
where $d^{2}z$ and $d^{2}\zeta$ are the respective Lebesgue measures on the
complex planes. The positive weight function $0\leq t\mapsto\varpi_{q}(t)$
solves the following Stieltjes moment problem:
\begin{equation}
\int_{0}^{\infty}t^{n}\varpi_{q}\left(  t\right)  \,dt=x_{n}!=n!\,q^{\dfrac
{n\left(  n+1\right)  }{2}}\,,\quad q\geq1\,, \label{momq}%
\end{equation}
where $x_{n}\overset{\mathrm{def}}{=}nq^{n}$ and we have adopted the
generalized factorial notation as $x_{n}!\overset{\mathrm{def}}{=}x_{n}%
x_{n-1}\cdots x_{1}$, $x_{0}!=1$.

In the Hilbert space%
\[
L^{2}\left(  \mathbb{C}^{2},\mu\left(  d\mathbf{x}\right)  \right)
=L^{2}\left(  \mathbb{C},e^{-|z|^{2}}\,\frac{d^{2}z}{\pi}\right)  \otimes
L^{2}\left(  \mathbb{C},\varpi_{q}(|\zeta|^{2})\,\frac{d^{2}\zeta}{\pi
}\right)
\]
we select the orthonormal set of functions
\begin{equation}
\Phi_{m,n}\left(  \mathbf{x}\right)  \overset{\mathrm{def}}{=}\frac{\bar
{z}^{m}}{\sqrt{m!}}\,\frac{\bar{\zeta}^{n}}{\sqrt{x_{n}!}}\, ,
\label{orthoset}%
\end{equation}
which we put in one-to-one correspondence with the elements $|m,n\rangle$,
$m,n\in\mathbb{N}$, of an orthonormal basis of a separable Hilbert space
$\mathcal{H}$. The states (\ref{orthoset}) obey a finite sum property for any
$\mathbf{x}\in\mathbb{C}^{2}$:
\begin{equation}
\sum_{m,n\in\mathbb{N}}\left\vert \Phi_{m,n}\left(  \mathbf{x}\right)
\right\vert ^{2}=e^{|z|^{2}}\,\mathcal{E}_{q}\left(  \left\vert \zeta
\right\vert ^{2}\right)  <\infty\,, \label{finsumprop}%
\end{equation}
where $\mathcal{E}_{q}\left(  t\right)  $ is the generalized \textquotedblleft
exponential\textquotedblright\ built from the sequence $x_{n}$:
\begin{equation}
\mathcal{E}_{q}\left(  t\right)  =\sum_{n=0}^{\infty}\frac{t^{n}}{x_{n}!}%
=\sum_{n=0}^{\infty}q^{-\frac{n\left(  n+1\right)  }{2}}\frac{t^{n}}{n!}\,.
\label{nzeta}%
\end{equation}
It is clear that, due to the condition $q\geq1$, the convergence radius of
this power series is infinite. On the other hand it is zero if $q<1$. In the
sequel we use the notation
\[
\mathcal{N}\left(  \mathbf{x}\right)  \overset{\mathrm{def}}{=}e^{\left\vert
z\right\vert ^{2}}\,\mathcal{E}_{q}\left(  \left\vert \zeta\right\vert
^{2}\right)  \,.
\]

We now introduce the CS corresponding to the above choice of orthonormal set.
They are elements of $\mathcal{H}$ defined by:
\begin{align}
\left\vert z,\zeta\right\rangle  &  \equiv|\mathbf{x}\rangle\overset
{\mathrm{def}}{=}\frac{1}{\sqrt{\mathcal{N}\left(  \mathbf{x}\right)  }}%
\sum_{m,n}\overline{\Phi_{m,n}\left(  \mathbf{x}\right)  }\,\left\vert
m,n\right\rangle \nonumber\\
&  =\frac{e^{-\frac{\left\vert z\right\vert ^{2}}{2}}}{\sqrt{\mathcal{E}%
_{q}\left(  \left\vert \zeta\right\vert ^{2}\right)  }}\sum_{m,n}\frac{z^{m}%
}{\sqrt{m!}}\,\frac{\zeta^{n}}{\sqrt{x_{n}!}}\left\vert m,n\right\rangle \,.
\label{csdef}%
\end{align}

By construction, these states are labeled by points of $\mathbb{C}^{2}$ and
have three important properties \cite{GazKl99}:

\begin{enumerate}
\item \textbf{Continuity}\newline The map $\mathbb{C}^{2}\ni\mathbf{x}%
=(z,\zeta)\mapsto\left\vert z,\zeta\right\rangle \in\mathcal{H}$ is continuous.

\item \textbf{Normalization}%
\begin{equation}
\left\langle z,\zeta\right.  \left\vert z,\zeta\right\rangle =1\,.
\label{normcs}%
\end{equation}

\item \textbf{Resolution of the unity}\newline The states form a continuous
overcomplete set in $\mathcal{H}$ which resolves the identity:
\begin{equation}
\int_{\mathbb{C}^{2}}\mu\left(  d\mathbf{x}\right)  \,\mathcal{N}\left(
\mathbf{x}\right)  \,\left\vert \mathbf{x}\right\rangle \left\langle
\mathbf{x}\right\vert =\mathbb{I}_{\mathcal{H}}\,. \label{resun}%
\end{equation}

\end{enumerate}

This resolution of the unity allows one to proceed to what we call CS
quantization or Berezin-Klauder-Toeplitz quantization. It consists in mapping
functions $f(\mathbf{x})$ (or even distributions) on $\mathbb{C}^{2}$ to
operators $A_{f}$ in $\mathcal{H}$ given as continued superpositions of CS
projectors weighted by $f$:
\begin{equation}
A_{f}\overset{\mathrm{def}}{=}\int_{\mathbb{C}^{2}}\mu\left(  d\mathbf{x}%
\right)  \,f\left(  \mathbf{x}\right)  \,\mathcal{N}\left(  \mathbf{x}\right)
\,\left\vert \mathbf{x}\right\rangle \left\langle \mathbf{x}\right\vert \,.
\label{CSquant}%
\end{equation}
A reasonable requirement on $f$ or on considered distributions is that the
mean value or \emph{lower symbol} of $A_{f}$ in CS, $\left\langle
\mathbf{x}\right\vert A_{f}\left\vert \mathbf{x}\right\rangle $, should be
smooth functions on $\mathbb{C}^{2}$.

In the specific above formulated problem of the particle in a magnetic field
(\ref{A.1}), the value of $q$ is fixed to $q=e^{2}\approx7.39$. The derivation
\cite{KowRe05} is based on the algebraic construction of CS for the motion of
a particle on a circle given in \cite{KowReP96}. Here we take the freedom to
consider \underline{any} value of $q$ in the range $q\in\lbrack1,\infty)$.

Our aim is to find the weight function $\varpi_{q}\left(  t\right)  $ solving
the Stieltjes moment problem (\ref{momq}) for any $q>1$ since for $q=1$ the
solution is well known, $\varpi_{1}\left(  t\right)  =e^{-t}$, which, for a
particle in a magnetic field, corresponds to the standard CS of Malkin and
Man'ko \cite{MalMa68}. For that purpose, we will use a Mellin inverse
transformation. But before, it is necessary to recall the conditions of
existence of a Mellin inverse transform in the present situation.

\begin{theorem}
(see \cite{mellinbook}) Suppose that a function $\Phi\left(  z\right)  $ of
the complex variable $z$, regular in the strip $S=\left\{  z=\sigma
+i\tau\,:\,a<\sigma<b\right\}  $, is such that $\Phi\left(  z\right)
\rightarrow0$ as $\left\vert \tau\right\vert \rightarrow\infty$ uniformly in
the strip $a+\eta\leq\sigma\leq b-\eta$ for any arbitrarily small $\eta>0$.
If
\begin{equation}
\int_{-\infty}^{+\infty}\left\vert \Phi\left(  \sigma+i\tau\right)
\right\vert \,d\tau<\infty\label{intconsmell}%
\end{equation}
for each $\sigma\in\left(  a,b\right)  $ and if a function $\phi\left(
x\right)  $ is defined by
\begin{equation}
\phi\left(  x\right)  =\frac{1}{2\pi i}\int_{c-i\infty}^{c+i\infty}%
x^{-z}\,\Phi\left(  z\right)  \,dz \label{con2smell}%
\end{equation}
for $x>0$ and a fixed $c\in\left(  a,b\right)  $, then we have the existence
of $\Phi\left(  z\right)  $ as the Mellin transform of $\phi\left(  x\right)
$, namely
\begin{equation}
\Phi\left(  z\right)  =\int_{0}^{+\infty}\phi\left(  x\right)  \,x^{z-1}%
\,dx\,. \label{melltrans}%
\end{equation}

\end{theorem}

In the case under consideration, we have, with $z=\sigma+i\tau$,
\[
\Phi\left(  z\right)  =\Gamma\left(  z\right)  \,q^{\frac{z\left(  z-1\right)
}{2}}=\Gamma\left(  \sigma+i\tau\right)  \,q^{\frac{\sigma\left(
\sigma-1\right)  }{2}}\,q^{i\tau\frac{2\sigma-1}{2}}\,q^{-\frac{\tau^{2}}{2}%
}\,.
\]
From the boundedness property of the Gamma function,
\[
\left\vert \Gamma\left(  \sigma+i\tau\right)  \right\vert =\left\vert \int
_{0}^{+\infty}t^{\sigma-1}\,t^{i\tau}\,e^{-t}\,dt\right\vert \leq\int
_{0}^{+\infty}t^{\sigma-1}\,e^{-t}\,dt=\Gamma\left(  \sigma\right)
<\infty\quad\forall\,\sigma>0\,,
\]
we infer the boundedness property for $\Phi$:
\[
\left\vert \Phi\left(  z\right)  \right\vert \leq\Gamma\left(  \sigma\right)
\,q^{\frac{\sigma\left(  \sigma-1\right)  }{2}}\,q^{-\frac{\tau^{2}}{2}}\,.
\]
Hence, we can assert that
\[
\left\vert \Phi\left(  z\right)  \right\vert \underset{\tau\rightarrow\infty
}{\rightarrow}0\,\ \text{for all}\ \sigma\in\left(  a,b\right)
\,\ \text{with}\ a\geq0\,.
\]
Moreover, the condition
\[
\int_{-\infty}^{+\infty}\left\vert \Phi\left(  \sigma+i\tau\right)
\right\vert \,d\tau\leq\Gamma\left(  \sigma\right)  \,q^{\frac{\sigma\left(
\sigma-1\right)  }{2}}\int_{-\infty}^{+\infty}q^{-\frac{\tau^{2}}{2}}%
\,d\tau=\sqrt{\frac{2\pi}{\ln{q}}}<\infty~,
\]
is clearly fulfilled for all $q>1$. Let us now attempt to determine the
function $\phi\left(  x\right)  $ formally defined for $x>0$ by
\[
\phi\left(  x\right)  =\frac{1}{2\pi}\int_{c-i\infty}^{c+i\infty}x^{-c-i\tau
}\,\Gamma\left(  c+i\tau\right)  \,q^{\frac{c\left(  c-1\right)  }{2}%
}\,q^{i\tau\frac{2c-1}{2}}\,q^{-\frac{\tau^{2}}{2}}\,d\tau\,.
\]
Choosing $c=1$ for simplicity purpose, and writing $q=e^{\lambda}$,
$\lambda>0$, we obtain the integral
\[
\phi\left(  x\right)  =\frac{1}{2\pi x}\int_{-\infty}^{+\infty}\left(
\frac{x}{\sqrt{q}}\right)  ^{-i\tau}\,\Gamma\left(  1+i\tau\right)
\,e^{-\lambda\frac{\tau^{2}}{2}}\,d\tau\,.
\]
It is legitimate to use here the integral representation of the Gamma
function, $\Gamma\left(  1+i\tau\right)  =\int_{0}^{+\infty}t^{i\tau}%
\,e^{-t}\,dt$ and to invert the order of integration from the Fubini theorem.
Hence,
\[
\phi\left(  x\right)  =\frac{1}{2\pi x}\int_{0}^{+\infty}dt\,e^{-t}%
\,\int_{-\infty}^{+\infty}\,e^{-iy\tau}\,e^{-\lambda\frac{\tau^{2}}{2}}%
\,d\tau\,,
\]
where we have introduced the variable $y\in\mathbb{R}$ such that
$x=t\,\sqrt{q}\,e^{y}$. Performing the integral on $\tau$ which is the Fourier
transform of a Gaussian, we obtain:
\[
\phi\left(  x\right)  =\frac{1}{\sqrt{2\pi\lambda}\,x}\int_{0}^{+\infty
}dt\,e^{-t}e^{-\frac{y^{2}}{2\lambda}}=\frac{1}{\sqrt{2\pi q\ln q}}\int
_{0}^{+\infty}du\,e^{-\frac{x}{\sqrt{q}}u}e^{-\frac{\left(  \ln u\right)
^{2}}{2\ln q}}\,,
\]
where we have introduced the new integration variable $u=\frac{\sqrt{q}}%
{x}\,t$.

We have eventually arrived at the following positive answer to our moment problem.

\begin{proposition}
The solution of the moment problem
\[
\int_{0}^{\infty}t^{n}\varpi_{q}\left(  t\right)  \,dt=n!\,q^{\dfrac{n\left(
n+1\right)  }{2}}~,\ q\geq1~,
\]
is given under the form of the following Laplace transform:
\begin{equation}
\varpi_{q}\left(  t\right)  =\frac{1}{\sqrt{2\pi q\ln q}}\int_{0}^{+\infty
}du\,e^{-\frac{t}{\sqrt{q}}u}e^{-\frac{\left(  \ln u\right)  ^{2}}{2\ln q}%
}=\frac{1}{\sqrt{2\pi q\ln q}}\,\mathcal{L}\left[  e^{-\frac{\left(  \ln
u\right)  ^{2}}{2\ln q}}\right]  \left(  \frac{t}{\sqrt{q}}\right)
\,.\nonumber
\end{equation}

\end{proposition}

\section{CS quantization of the complex plane}

Since from the very beginning we deal with a tensor product of two Hilbert
spaces and the resulting tensor products of standard Glauber-Sudarshan CS
\cite{Gla63,Sud63}\ ($q=1$) with non-standard CS ($q>1$), let us examine the
CS quantization of the complex plane produced by the latter, knowing that at
the limit $q\rightarrow1$ we get back the canonical quantization \cite{Ber65}.
Let us start with the family of CS
\begin{equation}
\mathbb{C}\ni\zeta\mapsto\left\vert \zeta\right\rangle =\frac{1}%
{\sqrt{\mathcal{E}_{q}\left(  \left\vert \zeta\right\vert ^{2}\right)  }}%
\sum_{n=0}^{+\infty}\frac{\zeta^{n}}{\sqrt{x_{n}!}}\left\vert n\right\rangle
\in\mathcal{K}\,,\quad x_{n}=n\,q^{n}\,,\,q\geq1~, \label{qcsdef}%
\end{equation}
where $\{\left\vert n\right\rangle \,,\,n\in\mathbb{N}\}$ is an orthonormal
basis of the separable Hilbert space $\mathcal{K}$, the latter being
arbitrarily chosen. The CS quantization of the complex plane with these states
is the map
\begin{equation}
f\left(  \zeta,\bar{\zeta}\right)  \mapsto\int_{\mathbb{C}}\frac{d^{2}\zeta
}{\pi}\,\varpi_{q}\left(  \left\vert \zeta\right\vert ^{2}\right)  \,f\left(
\zeta,\bar{\zeta}\right)  \,\mathcal{E}_{q}\left(  \left\vert \zeta\right\vert
^{2}\right)  \,\left\vert \zeta\right\rangle \left\langle \zeta\right\vert
\overset{\mathrm{def}}{=}\hat{f}\,, \label{qcsquant}%
\end{equation}
defining the operator $\hat{f}$ in $\mathcal{K}$. Note that the complex plane
is not equipped any more with its usual symplectic 2-form $\frac{i}{2\pi
}d\zeta\wedge d\bar{\zeta}$, but instead is equipped with the weighted 2-form
$\varpi( \vert\zeta\vert^{2} )\,\mathcal{E}_{q} ( \vert\zeta\vert^{2} )
\,\frac{i}{2\pi}d\zeta\wedge d\bar{\zeta}$.

For the simplest functions $f ( \zeta,\bar{\zeta} ) =\zeta$ and $f (
\zeta,\bar{\zeta}) =\bar{\zeta}$ we obtain lowering and raising operators
respectively,
\begin{align}
\widehat{\zeta}  &  =A\,,\quad A\,\left\vert n\right\rangle =\sqrt{x_{n}%
}\left\vert n-1\right\rangle \,,\quad A\,\left\vert 0\right\rangle
=0\,,\quad\text{lowering operator}~,\nonumber\\
\widehat{\bar{\zeta}}  &  =A^{\dag}\,,\quad A^{\dag}\,\left\vert
n\right\rangle =\sqrt{x_{n+1}}\left\vert n+1\right\rangle ~,\quad\text{raising
operator}\,. \label{stoper}%
\end{align}
These operators obey the non-canonical commutation rule: $\left[  A,A^{\dag
}\right]  =x_{N+1}-x_{N}$, where $x_{N}$ is defined by $x_{N}=A^{\dag}A$ and
is such that its spectrum is exactly $\left\{  x_{n}\,,\,n\in\mathbb{N}%
\right\}  $ with eigenvectors $x_{N}\left\vert n\right\rangle =x_{n}\left\vert
n\right\rangle $. The linear\ span of the triple $\left\{  A,A^{\dag}%
,x_{N}\right\}  $ is obviously not closed under commutation and the set of
resulting commutators gives generically rise to an infinite Lie algebra. The
fact that the complex plane has become non-commutative through this
quantization is apparent from the quantization of the real and imaginary parts
of $\zeta=\left(  q+ip\right)  /\sqrt{2}$:
\begin{equation}
\hat{q}=\frac{1}{\sqrt{2}}\left(  A+A^{\dag}\right)  \,,\ \hat{p}=\frac
{1}{\sqrt{2}i}\left(  A-A^{\dag}\right)  \,,\ \left[  \hat{q},\hat{p}\right]
=i\left(  x_{N+1}-x_{N}\right)  \,. \label{stQP}%
\end{equation}

Generally, given a function $f$ on the complex plane, the resulting operator
$\hat{f}$, if it exists, at least in a weak sense, acts on the Hilbert space
${\mathcal{K}}$ with orthonormal basis $\left\vert n\right\rangle $: the
integral
\begin{equation}
\left\langle \psi\right\vert \hat{f}\left\vert \psi\right\rangle
=\int_{\mathbb{C}}\frac{d^{2}\zeta}{\pi}\,\varpi_{q}(\left\vert \zeta
\right\vert ^{2})\,f\left(  \zeta,\bar{\zeta}\right)  \,\mathcal{E}_{q}\left(
\left\vert \zeta\right\vert ^{2}\right)  \,\left\vert \left\langle
\psi\right\vert \left.  \zeta\right\rangle \right\vert ^{2}~, \label{wksssymb}%
\end{equation}
should be finite for any $\psi\in$ some dense subset in $\mathcal{K}$. One
should notice that if $\psi$ is normalized then (\ref{wksssymb}) represents
the mean value of the function $f$ with respect to the $\psi$-dependent
probability distribution $\zeta\mapsto\left\vert \left\langle \psi\right\vert
\left.  \zeta\right\rangle \right\vert ^{2}$ on the complex plane.

In order to be mostly rigorous on this important point, let us adopt the
following acceptance criteria for a function to belong to the class of
quantizable classical observables.

\begin{definition}
A function $\mathbb{C}\ni\zeta\mapsto f\left(  \zeta,\bar{\zeta}\right)
\in\mathbb{C}$ is a \emph{CS quantizable classical observable} along the map
$f\mapsto\hat{f}$ defined by (\ref{qcsquant}) if the map $\mathbb{C}\ni
\zeta=\frac{1}{2}\left(  q+ip\right)  \equiv\left(  q,p\right)  \mapsto
\left\langle \zeta\right\vert \hat{f}\left\vert \zeta\right\rangle $ is a
smooth ($\sim\in C^{\infty}$) function with respect to the $\left(
q,p\right)  $ coordinates of the complex plane.
\end{definition}

The function $f$ is the \emph{upper} \cite{LieFeK94} or \emph{contravariant}
\cite{Ber75} symbol of the operator $\hat{f}$, and $\left\langle
\zeta\right\vert \hat{f}\left\vert \zeta\right\rangle $ is the \emph{lower}
\cite{LieFeK94} or \emph{covariant} \cite{Ber75} symbol of the operator
$\hat{f}$.

In \cite{ChaGaY08} such a definition is extended to distributions with compact
support on the plane.

Hence, localization properties in the complex plane from the point of view of
the sequence $\{ x_{n} \}_{n\in\mathbb{N}}$ should be examined from the shape
(versus $\zeta$) of the respective lower symbols%

\begin{equation}
\check{q}(z)\overset{\mathrm{def}}{=}\left\langle \zeta\right\vert \hat
{q}\left\vert \zeta\right\rangle \,,\check{p}(\zeta)\overset{\mathrm{def}}%
{=}\left\langle \zeta\right\vert \hat{p}\left\vert \zeta\right\rangle \,,
\label{lowsymbQP}%
\end{equation}
and the ``noncommutative reading\textquotedblright\ of the complex plane
should be encoded in the behavior of the lower symbol $\left\langle
\zeta\right\vert \left[  \hat{q},\hat{p}\right]  \left\vert \zeta\right\rangle
$ of the commutator $\left[  \hat{q},\hat{p}\right]  $. The study, within the
above framework, of the product of dispersions
\begin{equation}
\left(  \Delta_{\zeta}\hat{q}\right)  \,\left(  \Delta_{\zeta}\hat{p}\right)
=\frac{1}{2}\left\langle \zeta\right\vert \left(  {x}_{N+1}-{x}_{N}\right)
\left\vert \zeta\right\rangle \,, \label{uncert}%
\end{equation}
expressed in states $|\zeta\rangle$ should be thus relevant.

Let us now consider the CS quantized version of the classical harmonic
oscillator Hamiltonian $H=\frac{1}{2}\left(  p^{2}+q^{2}\right)  =\left\vert
\zeta\right\vert ^{2}$. We get the diagonal operator $\hat{H}=A\,A^{\dag}$
with spectrum $x_{N+1}$. It is then natural to investigate the time evolution
of the quantized version of the classical phase space point through its lower
symbol:
\begin{equation}
\check{\zeta}\left(  t\right)  \overset{\mathrm{def}}{=}\left\langle
\zeta\right\vert e^{-i\hat{H}t}\hat{\zeta}e^{i\hat{H}t}\left\vert
\zeta\right\rangle =\zeta\,\frac{1}{\mathcal{E}_{q}\left(  \left\vert
\zeta\right\vert ^{2}\right)  }\,\sum_{n=0}^{+\infty}\frac{\left\vert
\zeta\right\vert ^{2n}}{x_{n}!}\,e^{-i\left(  x_{n+2}-x_{n+1}\right)  t}\,,
\label{timeev}%
\end{equation}
and to compare with the phase space circular classical trajectories, at
different values of $q$. This study will be carried out in a forthcoming paper
devoted to the noncommutive version of the model considered in this paper.

\section{Final remarks}

We have generalized the construction of coherent states for a particle in a
magnetic field proposed in \cite{KowRe05}. We have proved the resolution of
the identity for the states by explicitly solving a specific version of the
mathematical Stieltjes moment problem. So, using the CS, we are in position to
fulfill the Berezin-Klauder-Toeplitz quantization.

As an application of this construction, in a future work, we will present the
CS quantization of a charged particle in a uniform magnetic field in a
non-commutative geometry. In addition we will investigate the dispersion
relations and the time evolution of the CS quantized version of the system for
different members of our family of CS, i.e., at different values of our
parameter $q$.

\begin{acknowledgement}
M.C.B. thanks FAPESP; D.M.G. thanks FAPESP and CNPq for permanent support.
This work was financed by CAPES-COFECUB, N%
${{}^\circ}$
PH 566/07.
\end{acknowledgement}


\begin{thebibliography}{99}                                                                                               %


\bibitem {ab1}J.R. Klauder, E.C. Sudarshan, \emph{Fundamentals of Quantum
Optics}, Benjamin, 1968.

\bibitem {ab2}I.A. Malkin, V.I. Man'ko, \emph{Dynamical Symmetries and
Coherent States of Quantum Systems}, Nauka, Moscow, 1979, pp.320.

\bibitem {ab3}I.R. Klauder, B.S. Skagerstam, \emph{Coherent States,
Applications in Physics and Mathematical Physics}, World Scientific,
Singapore, 1985, pp.991.

\bibitem {ab5}V.G. Bagrov, D.M. Gitman, \emph{Exact Solutions of Relativistic
Wave Equations}, Kluwer acad. publish., 1990, pp.323.

\bibitem {ab4}A.M. Perelomov, \emph{Generalized Coherent States and Their
Applications}, Springer-Verlag, 1986.

\bibitem {aag}S.T. Ali, J.P. Antoine, and J.P. Gazeau, \emph{Coherent states,
wavelets and their generalizations. Graduate Texts in Contemporary Physics},
Springer-Verlag, New York, 2000.

\bibitem {ab6}A.M. Perelomov, Comm. Math. Phys. \textbf{26} (1972) 222.

\bibitem {ab7}M. Rasetti, Int. J. Theor. Phys. \textbf{13} (1975) 425.

\bibitem {ab8}F.A. Berezin, Comm. Math. Phys. \textbf{40} (1975) 153.

\bibitem {ab9}F.A. Berezin, M.A. Shubin, \emph{Schrodinger Equation}, Moscow
State University, Moscow, 1983.

\bibitem {KowRe05}K. Kowalski and J. Rembielinski, J. Phys. A \textbf{38}
(2005) 8247.

\bibitem {KowReP96}K. Kowalski, J. Rembielinsk, and L.C. Papaloucas, J. Phys.
A \textbf{29} (1996) 4149.

\bibitem {MalMa68}I.A. Malkin and V.I. Man'ko, Zh. Eksp. Teor. Fiz.
\textbf{55} (1968) 1014 [Sov. Phys. - JETP \textbf{28}, no.3 (1969) 527].

\bibitem {GazKl99}J.P. Gazeau and J.R. Klauder, J. Phys. A \textbf{32} (1999) 123.

\bibitem {mellinbook}B. Davies, \emph{Integral Transforms and their
Applications}, Springer; 3rd edition, 2002.

\bibitem {Gla63}R.J. Glauber, Phys. Rev. Lett. \textbf{10} (1963) 84.

\bibitem {Sud63}E.C.G. Sudarshan, Phys. Rev. Lett. \textbf{10} (1963) 277.

\bibitem {Ber65}F.A. Berezin, \emph{The Method of Second Quantization}, Nauka,
Moscow, 1965.

\bibitem {LieFeK94}E. Lieb, D. H. Feng, J. R. Klauder, and M. Strayer (eds.),
\emph{Coherent States: Past, Present and Future}, World Scientific, Singapore, 1994.

\bibitem {Ber75}F.A. Berezin, Commun. Math.Phys. \textbf{40} (1975) 153.

\bibitem {ChaGaY08}B. Chakraborty, J.P. Gazeau, and A. Youssef, arXiv:0805.1847v1.
\end{thebibliography}
\end{document}